\documentclass[pdflatex,sn-mathphys-num]{sn-jnl}% 
\usepackage{graphicx}%
\usepackage{multirow}%
\usepackage{amsmath,amssymb,amsfonts}%
\usepackage{amsthm}%
\usepackage{mathrsfs}%
\usepackage[title]{appendix}%
\usepackage{xcolor}%
\usepackage{textcomp}%
\usepackage{manyfoot}%
\usepackage{booktabs}%
\usepackage{algorithm}%
\usepackage{algorithmicx}%
\usepackage{algpseudocode}%
\usepackage{listings}%

\usepackage{bm, float}
\usepackage{tabularx}
\usepackage{booktabs}  
\usepackage{graphicx}
\usepackage{array}    
\usepackage{mathtools}
\usepackage{makecell}
\usepackage{pifont}
\usepackage{xcolor}
\newcommand{\cmark}{\textcolor{green!60!black}{\ding{51}}} % tick xanh
\newcommand{\xmark}{\textcolor{red}{\ding{55}}}            % x đỏ

\theoremstyle{thmstyleone}%
\theoremstyle{thmstyletwo}%

\theoremstyle{thmstylethree}%

\begin{document}

\title[Towards Automated Selection of Quantum Encoding Circuits via Meta-Learning]{Towards Automated Selection of Quantum Encoding Circuits via Meta-Learning}


\author[1,2]{\fnm{Dao Duy} \sur{Tung}}

\author*[3]{\fnm{Nguyen Quoc} \sur{Chuong}}
\email{nguyenquocchuong2@duytan.edu.vn}

\author[4,2]{\fnm{Vu Tuan} \sur{Hai}}
\email{haivt@uit.edu.vn}

\author[5,6]{\fnm{Le Bin} \sur{Ho}}
\email{binho@fris.tohoku.ac.jp}

\author[1,2]{\fnm{Lan Nguyen} \sur{Tran}}
\email{tnlan@hcmus.edu.vn}

% Affiliations
\affil[1]{\orgname{University of Science, Vietnam National University, Ho Chi Minh City}, \country{Vietnam}}

\affil[2]{\orgname{Vietnam National University, Ho Chi Minh City}, \country{Vietnam}}

\affil[3]{\orgname{Institute of Fundamental and Applied Sciences, Duy Tan University, Ho Chi Minh City}, \country{Vietnam}}

\affil[4]{\orgname{University of Information Technology, Vietnam National University, Ho Chi Minh City}, \country{Vietnam}}

\affil[5]{\orgname{Graduate School of Engineering, Tohoku University}, \city{Sendai}, \postcode{980-8579}, \country{Japan}}

\affil[6]{\orgname{Frontier Research Institute for Interdisciplinary Sciences, Tohoku University}, \city{Sendai}, \postcode{980-8578}, \country{Japan}}


\abstract{In recent years, quantum kernel methods have shown promising applications on near-term quantum devices. However, selecting an appropriate encoding circuit for a given dataset requires costly evaluation of multiple candidates, formulated as a meta-learning problem. In this paper, we propose an automated recommender that utilizes the intrinsic characteristics of datasets to predict the optimal circuit without any quantum evaluation. Nine candidates are assessed alongside 24 classical complexity metrics serving as features, evaluated through two training approaches with four configurations, along with 14 machine learning models. Both approaches achieve Top-$3$ accuracy of up to 85.7\% in identifying the best-performing encoding circuit, and demonstrate that classical data complexity metrics provide sufficient predictive signal for circuit selection.}


\maketitle

\section{Introduction}

As the amount of data continues to grow, classical machine learning is approaching its computational limits \cite{rspa}, particularly for tasks involving high-dimensional feature spaces, complex nonlinear decision boundaries, and large-scale optimization problems~\cite{bishop2016pattern, doi:10.1137/16M1080173}. Quantum computing offers a new computational paradigm based on the principles of quantum mechanics, with the potential to provide speedups for certain classes of problems~\cite{doi:10.1137/S0097539795293172, 10.1145/237814.237866}. This has led to the emergence of quantum machine learning (QML), which aims to leverage quantum resources to enhance and extend classical learning methods~\cite{adcock2015advancesquantummachinelearning, Biamonte2017}.

Among QML approaches, quantum kernel methods (QKMs) have attracted considerable interest, particularly since~\cite{Liu2021} demonstrated a rigorous quantum advantage for a classification task. In this framework, an encoding circuit maps each classical input into a quantum state, and the kernel value is given by the fidelity (overlap) between pairs of encoded states. Since the kernel is determined by the choice of encoding circuit, different circuits lead to different quantum kernels and thus different similarity structures over the data. Unlike variational approaches such as quantum neural networks~\cite{Abbas2021, Pan2023, 10.1145/3529756} and the variational quantum eigensolver~\cite{Peruzzo2014}, which require iterative optimization of circuit parameters and can suffer from barren plateaus~\cite{Larocca2025, Cerezo2021}, the encoding circuit in QKMs remains fixed during training, and the learning task is carried out by a classical algorithm using the resulting kernel matrix~\cite{PhysRevLett.122.040504}. This design eliminates exposure to barren plateaus, yields a convex training objective with a guaranteed global optimum, and ensures that the optimal model is always recoverable via kernel-based training~\cite{schuld2021supervisedquantummachinelearning}.

Despite these advantages, choosing a suitable encoding circuit for a given problem remains challenging. Circuits with high expressibility often lead to an exponential concentration of kernel values, making different data points difficult to distinguish without a very large number of samples. In addition, highly expressive circuits may not match the structure of the target problem, resulting in poor learning performance. From a theoretical perspective, this is reflected in the Mercer decomposition, where many eigenvalues become exponentially small, rendering most components effectively unlearnable~\cite{10812182}. Moreover, the number of possible encoding circuits grows combinatorially with the number of qubits and gates, making a systematic search computationally intractable~\cite{neto2025datacomplexitymeasuresquantum}.

Several efforts have been made to automate quantum architecture selection. Ref.~\cite{10812182} proposed a method to discover encoding circuits using combinatorial optimization. However, this approach requires a costly optimization loop for each new dataset and does not explicitly incorporate the characteristics of the data. The Ref.~\cite{neto2025datacomplexitymeasuresquantum} introduced a strategy based on classical data complexity measures to recommend suitable architectures. Their method, however, focuses on variational quantum circuits rather than quantum kernel methods. So far, existing approaches do not directly address the problem of selecting encoding circuits for quantum kernel methods based on dataset characteristics.

In this work, we develop a framework to automatically select suitable encoding circuits for a given dataset. Our approach begins by characterizing the dataset using classical data complexity measures. These measures are computed with problexity~\cite{komorniczak2023problexity, lorena2018complex} and Qsun~\cite{Nguyen_2022} libraries, and capture key structural properties of the data. To evaluate the proposed framework, we use two validation strategies: Majority Voting (MV) \cite{667881},  and Leave-One-Out Cross-Validation (LOOCV) \cite{10.5555/1643031.1643047}, each combined with 14 classical machine learning classifiers.

The main contributions of this work are as follows:
\begin{itemize}
\item We formulate encoding circuit selection as a meta-learning problem.
\item We perform a comprehensive evaluation under multiple configurations.
\item We demonstrate that the proposed framework achieves Top-$3$ accuracy of up to 85.7\% on the meta-dataset while reducing computational cost up to 78\% compared to exhaustive quantum evaluation.
\end{itemize}

The remainder of this paper is organized as follows. Section~\ref{sec:background} introduces the theoretical background, covering complexity metrics and QKMs. Section~\ref{sec:method} describes the proposed framework, including the feature extraction pipeline, kernel evaluation procedure, and meta-learning formulation. Section~\ref{sec:experiment_results} presents the experimental results and analysis. Finally, Section~\ref{sec:conclusion} concludes with a discussion of key findings and future directions. Classical descriptors, existing encoding circuits, and classifiers are described in detail in the Appendix.

\section{Theoretical background}
\label{sec:background}

\subsection{Data complexity metrics}

Data complexity metrics describe the intrinsic characteristic of classification tasks through geometric and statistical properties, independent of any specific learning algorithm~\cite{lorena2018complex}. Early works emphasized the role of class geometry in feature space, introducing measures of class separability and methods to detect linear separability~\cite{990132}. Building on this idea, Ref.~\cite{22} proposed a meta-learning approach in which these measures are used as meta-features to predict the most suitable classifier for a given dataset. More recently, this approach has been extended to the quantum setting, showing that data complexity can also guide the choice of circuit architectures, including circuit structure and number of layers~\cite{neto2025datacomplexitymeasuresquantum}.

\subsection{Quantum kernel methods}

The kernel methods are a fundamental tool in both classical and quantum machine learning. A kernel function $K: \mathcal{X} \times \mathcal{X} \rightarrow \mathbb{R}$ measures similarity between data points without explicitly working in a high-dimensional space. By Mercer’s theorem~\cite{10.1098/rspa.1909.0075}, any valid kernel can be written as an inner product in a feature space, $K(x, x') = \langle \phi(x), \phi(x') \rangle_{\mathcal{F}}$, where $\phi: \mathcal{X} \rightarrow \mathcal{F}$ is a feature map and $\mathcal{F}$ denotes the feature space. This enables the \textit{kernel trick}: computing inner products through $K$ without constructing $\phi$~\cite{PhysRevLett.122.040504, schuld2021supervisedquantummachinelearning}.
With a nonlinear feature map, data that are not linearly separable in $\mathcal{X}$ may become separable in $\mathcal{F}$, as used in Support Vector Machines (SVMs)~\cite{10.1145/130385.130401}.

This framework can be naturally extended to the quantum setting. A unitary $U$ encodes the classical input $\bm{x}$ into a quantum state $|\phi(\bm{x})\rangle = U|0\rangle^{\otimes n}$, acting as a feature map in an $n$-qubit Hilbert space. The quantum kernel is defined as the squared fidelity between two encoded states:

\begin{align}\label{eq:fidelity}
K(\bm{x}, \bm{x}') = |\langle \phi(\bm{x})|\phi(\bm{x}')\rangle|^2.
\end{align}

For mixed states, Equation~\ref{eq:fidelity} becomes the Hilbert-Schmidt inner product, $K(\bm{x}, \bm{x}') = \mathrm{Tr}[\rho(\bm{x}')^\dagger \rho(\bm{x})]$ \cite{PhysRevA.106.042431}. The resulting kernel (Gram) matrix is then used by a classical model - forming a hybrid quantum-classical pipeline.

The quantum kernel is fully determined by $U$, so circuit selection is a key step for any learning task. A suitable encoding circuit maps similar inputs to similar quantum states, leading to a Gram matrix with clear separation between classes. In contrast, a poor encoding circuit may mix different classes or spread all data points too evenly. In practice, many existing works use a fixed circuit or test all possible encoding circuits by brute force, which becomes inefficient as the number of encoding circuits increases \cite{Huang2021}. This limitation motivates the need for an automated recommender that recommends the most suitable encoding circuit based on simple properties of the dataset.

\section{Proposed framework}
\label{sec:method}

\begin{figure*}[t]
    \centering
    \includegraphics[width=1.2\textwidth]{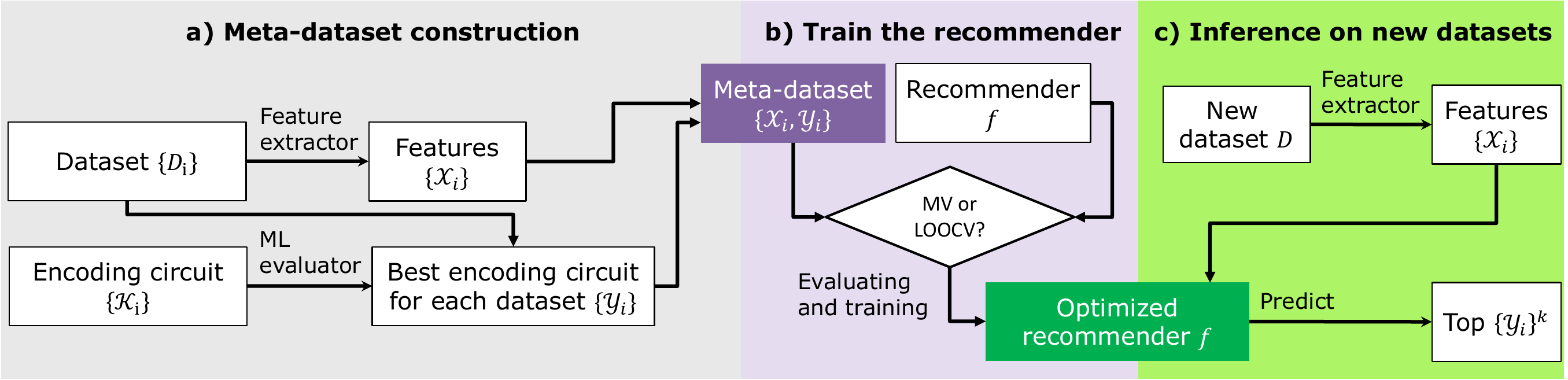}
    \caption{Overview of the proposed encoding circuit selection framework. Stage~(a): data processing and ground-truth label generation via ML evaluator. Stage~(b): Recommender training with multiple configurations. Stage~(c): $f$ recommend Top-$k$ encoding circuits as final output. 
    }
    \label{fig:scheme}
\end{figure*}

\begin{figure}[t]
    \centering
    \includegraphics[width=0.8\textwidth]{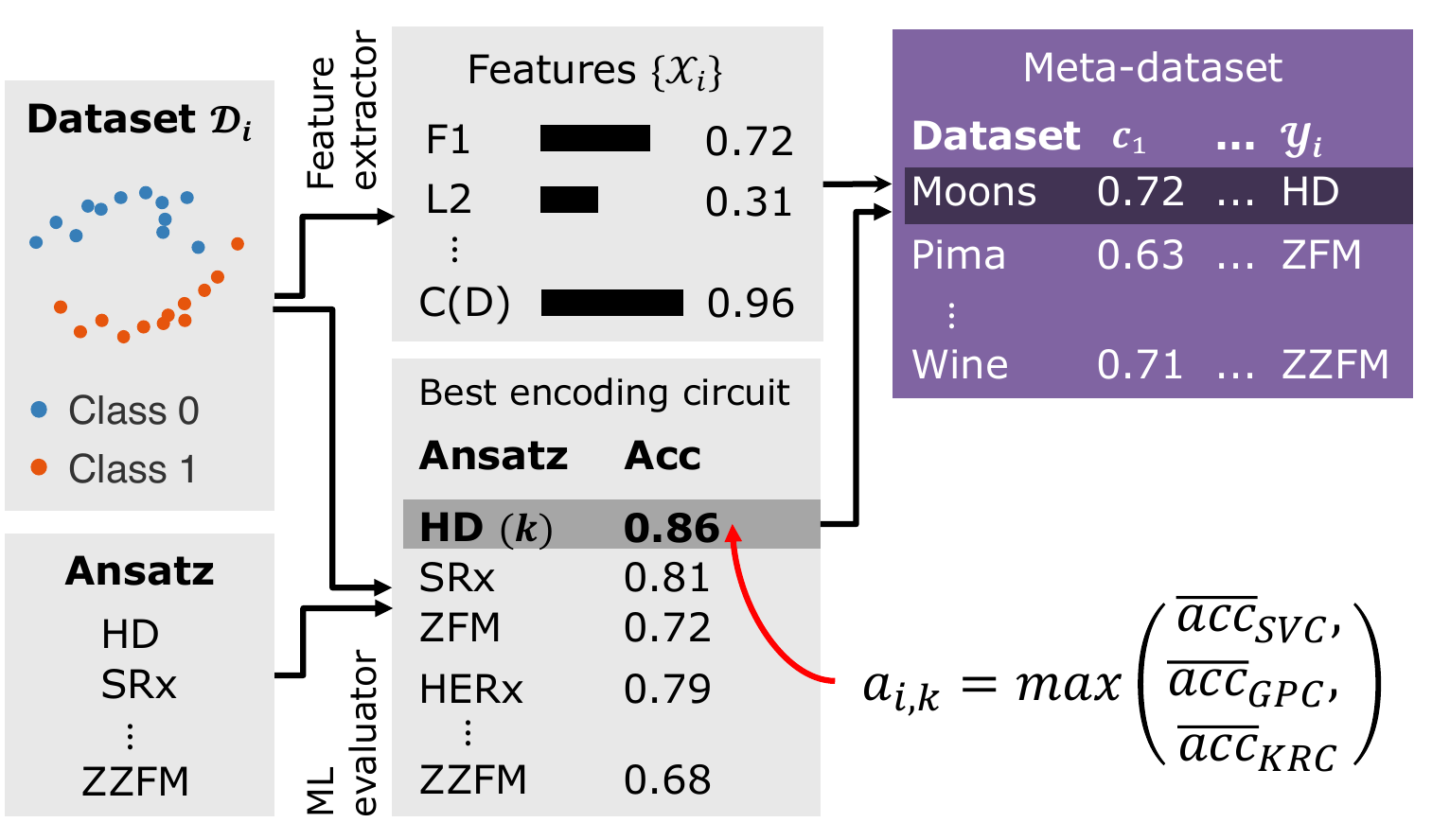}
    \caption{Illustration of the Moon dataset construction process in \textbf{ALL-IN} mode. We can use three classifiers: Support Vector Classifier (SVC)~\cite{Cortes1995}, Gaussian Process Classifier (GPC)~\cite{10.7551/mitpress/3206.001.0001}, and Kernel Ridge Classifier (KRC)~\cite{10.5555/645527.657464} to get the final accuracy $acc_{i,k}$. The symbol $\overline{\mathrm{acc}}$ denotes the mean test accuracy. }
    \label{fig:example}
\end{figure}

We formulate the recommender optimization task as a supervised meta-learning problem,
illustrated in Figure~\ref{fig:scheme}.
The framework operates on a collection of classification datasets and a set of candidate encoding circuits. As a classical machine learning problem, we separate the framework into 3 stages: (a) meta-dataset construction, (b) train the recommender, and (c) inference on new datasets.

\subsection{Meta-dataset construction}
\label{sec:meta-dataset}

For a list of dataset $\{\mathcal{D}_i\}$ and candidate encoding circuits
$\{\mathcal{K}_i\}$, two parallel processes
are carried out. First, the feature extractor returns a metric vector (called a feature)
$\mathcal{X}_i\in \mathbb{R}^d$ for each dataset $\mathcal{D}_i$, following classical descriptors in Table~\ref{tab:complexity_metrics}. Each circuits are then evaluated using different classifiers via the ML evaluator. We define circuit performance as the best achievable accuracy across classifiers, approximating an upper bound on circuit capability, as illustrated in Figure~\ref{fig:example}. 

The meta-dataset is constructed from $200$ binary classification datasets, comprising $174$ synthetic and $26$ real-world datasets. The synthetic portion is generated using eight generators - including standard ones from scikit-learn~\cite{scikit-learn} (Blobs, Circles, Moons) and custom implementations (XOR, Spiral, Checkerboard, Concentric Rings) - each instantiated with multiple configurations by varying hyperparameters. Each configuration is repeated across three sample sizes $\{100, 150, 200\}$. For real-world data, they are sampled at sizes $\{80, 100, 120, 150\}$. Although synthetic datasets are generated from common families, parameter variations introduce diverse decision boundaries. A detailed breakdown is provided in Table~\ref{tab:datasets}. 

\begin{table*}[t]
    \centering
    \caption{Description of datasets used in experiments.}
    \label{tab:datasets}
    
    \begin{tabular}{lcccc} % Thay đổi số lượng cột ở đây
        \toprule
        \textbf{Dataset} & \textbf{\#Features} & \textbf{\#Classes} & \textbf{\#Configs} & \textbf{Count} \\
        \midrule
        \multicolumn{5}{l}{\textit{Synthetic datasets (scikit-learn \& custom implementations)}} \\
        Blobs             & 2--4 & 2--4 & 14 & 42 \\
        Circles           & 2    & 2    & 8  & 24 \\
        Moons             & 2    & 2    & 7  & 21 \\
        Concentric Rings  & 2    & 2    & 8  & 24 \\
        XOR               & 2    & 2    & 6  & 18 \\
        Spiral            & 2    & 2    & 6  & 18 \\
        Checkerboard      & 2    & 2    & 9  & 27 \\
        \midrule
        \multicolumn{5}{l}{\textit{Real-world datasets (scikit-learn, UCI \& Kaggle)}} \\
        Iris~\cite{scikit-learn}                & 4  & 2 & 2 & 2 \\
        Wine~\cite{scikit-learn}                & 13 & 2 & 4 & 4 \\
        Breast Cancer~\cite{scikit-learn}       & 30 & 2 & 5 & 5 \\
        Pima Diabetes~\cite{Smith1988UsingTA}   & 8  & 2 & 5 & 5 \\
        Banknote ~\cite{banknote_authentication_267} & 4  & 2 & 5 & 5 \\
        Haberman~\cite{habermans_survival_43}  & 3  & 2 & 5 & 5 \\
        \midrule
        \textbf{Total}    &      &   &   & \textbf{200} \\
        \bottomrule
    \end{tabular}
\end{table*}

Each dataset undergoes a standardized pre-processing pipeline to ensure consistency across all experiments. First, features $\mathcal{X}$ are scaled in the range $[0, 1]$ using \texttt{MinMaxScaler} as problexity library requires normalized inputs. When the input dimensionality exceeds the available \#Qubits, Principal Component Analysis (PCA) is applied to reduce the feature space to \#Qubits-dimensional. Subsequently, the scaled values are remapped to $[0, \pi]$ for angle encoding, where each value corresponds to a rotation gate's parameter. Finally, the dataset is partitioned using an $8/2$ train-test strategy. 

\subsubsection{Feature extractor} has two extraction modes: the \textbf{SINGLE-IN}, in which only one metric is used; and the \textbf{ALL-IN}, in which all metrics are concatenated. Comparing these two configurations reveals whether a single informative metric suffices or whether the full feature set introduces redundancy, given the limited training size.

\subsubsection{ML evaluator} generates the label $\mathcal{Y}_i$, which indicates the Top-$k$ suitable encoding circuits (note that $k=|\mathcal{Y}_i|$). For example, $\mathcal{Y}_i=\{y_1,y_2\}=\{1,4\}$ means that $f$ recommends the $1^{\text{st}}$ and $4^{\text{th}}$ circuits in existing list. The ML evaluator has two labeling modes, $k=1 $ (\textbf{SINGLE-BEST-OUT}) - we predict the single circuit that achieves the highest test accuracy. A prediction is considered correct only if it exactly matches the best circuit. Otherwise, if $k>1$ (\textbf{TIED-BEST-OUT}), ML evaluator returns $k$ circuits within a tolerance threshold $\epsilon$, (e.g, within 1\% of the best accuracy). This formulation acknowledges that multiple encoding circuits often achieve statistically equivalent performance. In case $k=1$, it exhibits significant class imbalance, while $k>1$ provides a more balanced distribution. 

A natural question arises: ``\textit{Is selecting the single encoding circuit with the highest accuracy truly the most reliable criterion?}'' We adopt the \textbf{TIED-BEST-OUT} labeling mode for two reasons. First, the accuracy rankings among encoding circuits are sensitive to the choice of train–test split. In many datasets, several encoding circuits achieve accuracies within a narrow margin, making it unreliable to designate a single optimum. Second, recommending a set of equally competitive encoding circuits allows practitioners to select among them based on secondary criteria such as circuit depth, gate count, or hardware compatibility.

Because both the feature extractor and ML evaluator have two modes: \textbf{SINGLE-IN}, \textbf{ALL-IN}, and \textbf{SINGLE-BEST-OUT}, \textbf{TIED-BEST-OUT}. We have a total of four configurations that affect the training stage. All configurations are surveyed in Section~\ref{sec:experiment_results}.

\subsection{Train the recommender}
\label{sec:training}

After forming
$\mathcal{D}_{\text{meta}}=\{(\mathcal{X}_i, \mathcal{Y}_i)\}$ from Stage (a). The final goal is to train the recommender $f$ to recommend suitable encoding circuits for the dataset $\mathcal{D}_i$. i.e, $f:\mathcal{X}_i\rightarrow\mathcal{Y}_i$. We investigate two training strategies: MV (Algorithm~\ref{alg:mv}) and LOOCV (Algorithm~\ref{alg:loocv}). MV uses all classifiers simultaneously, which reduces sensitivity to any individual classifier's errors. In contrast, LOOCV evaluates every classifier, then selects one classifier that achieves the highest accuracy.

\subsubsection{Majority Voting (MV)}

Given an ensemble of $H$ trainable classifiers $\{h_1(.), \dots, h_H(.)\}$, each classifier is trained independently. Then, the final prediction for a feature $\mathcal{X}$, which returns only one circuit:
\begin{align}\label{eq:mv}
    \hat{y} &= \arg\max_{y} \sum_{j=1}^{H} 
    \mathbf{1}\left[h_j(\bm w_j)(\mathcal{X})= y\right]\nonumber\\
    &:=\text{MajorityVoting}(\{h_j(\bm w_j)\},\mathcal{X}),
\end{align}
where $\mathbf{1}[\cdot]$ is the indicator function. Although MV produces a single prediction, the vote distribution is used to derive Top-k recommendations

\begin{algorithm}[t]
\caption{MV evaluation and training}
\label{alg:mv}
\begin{algorithmic}
\Require Meta-dataset $\mathcal{D}_{\text{meta}}$, number of runs $R$
\Ensure Optimized recommender $f^{\text{MV}}$

\Statex \textit{// Evaluation phase}
\For{$r = 1$ to $R$}
    \State correct $\gets 0$
    \State $\mathcal{D}_{\text{train}}^{(r)}, \mathcal{D}_{\text{test}}^{(r)}\gets\text{SplitWithSeed}(\mathcal{D}_{\text{meta}}, r)$
    \State $\{h_j(\bm w_j^*)\}\gets$ Train($\{h_j(\bm w)\},\mathcal{D}_{\text{train}}^{(r)})$
    \For{each $(\mathcal{X}_i, \mathcal{Y}_i) \in \mathcal{D}_{\text{test}}^{(r)}$}
        \State $\hat{y} \gets \text{MajorityVoting}(\{h_j(\bm w_j^*)\},\mathcal{X})$ \Comment{see Equation~\ref{eq:mv}}
        \If{$\hat{y} \in \mathcal{Y}_i$}
            \State correct $\gets$ correct $+\ 1$
        \EndIf
    \EndFor
    \State $\text{acc}^{(r)} \gets \text{correct}/|\mathcal{D}_{\text{test}}^{(r)}|$
\EndFor
\State $\overline{\text{acc}} \gets \mathbb{E}[\text{acc}^{(r)}]$

\Statex \textit{// Training phase}
\State $f^{\text{MV}}\gets$ Train($\{h_j(\bm w_j)\},\mathcal{D}_{\text{meta}}$)
\State \Return $f^{\text{MV}}=\{h_1(\bm w_1^*),\ldots,h_H(\bm w^*_H)\}$
\end{algorithmic}
\end{algorithm}

\subsubsection{Leave-One-Out Cross-Validation (LOOCV)}

Given a training set of $N$ dataset instances, LOOCV iteratively holds out one instance as the test sample while training on the remaining $N - 1$ instances. This process is repeated exhaustively, and the final accuracy is computed as the proportion of correct predictions over $N$. In contrast with MV, the goal of LOOCV is to find the best classifier and eliminate the others.

\begin{algorithm}[t]
\caption{LOOCV evaluation and training}
\label{alg:loocv}
\begin{algorithmic}
\Require Meta-dataset $\mathcal{D}_{\text{meta}}$
\Ensure Optimized recommender $f^{\text{LOOCV}}$

\Statex \textit{// Evaluation phase}
\For{each $(h(.),m) \in \{h_1(.),\ldots,h_H(.)\} \times \{\textbf{SINGLE-IN},\textbf{ALL-IN}\}$}
    \State correct $\gets 0$
    \For{$i = 1$ to $N$}
        \State $\mathcal{D}_{\text{train}} \gets \mathcal{D}_{\text{meta}} \setminus (\mathcal{X}_i, \mathcal{Y}_i)$
        
        \State $\mathcal{D}_{\text{test}} \gets(\mathcal{X}_i, \mathcal{Y}_i)$
        
        \State $h(\bm w^*)\gets \text{Train}(\mathcal{D}_{\text{train}},m)$
        \State $\hat{y} \gets h(\bm w^*)(\mathcal{X}_i)$ \Comment{$h$ predicts}
        \If{$\hat{y} \in \mathcal{Y}_i$}
            \State correct $\gets$ correct $+\ 1$
        \EndIf
    \EndFor
    \State $\text{acc}(m, h) \gets \text{correct}/N$
\EndFor
\State $(h^*, m^*) \gets \arg\max_{(h, m)} \{\text{acc}(h, m)\}$

\Statex \textit{// Training phase}
\State $h^*(\bm w^*)\gets\text{Train}(\mathcal{D}_{\text{meta}},m^*)$
\State \Return $f^{\text{LOOCV}}=h^*(\bm w^*)$
\end{algorithmic}
\end{algorithm}

\subsection{Inference on new datasets}
\label{sec:inference}

Given an unseen dataset, only the feature extractor operates to produce a feature $ \mathcal{X}$. The recommender $f$ returns a vote distribution, and the Top-$k$ candidates are selected based on descending vote counts:

\begin{align}
    & \hat{\mathcal{Y}}=\text{Top-}k(\mathcal{X}) \coloneq \{\hat{y}_1, \ldots, \hat{y}_k\} \\
    &\text{s.t.} \quad \text{votes}[\hat{y}_1] \geq \cdots \geq \text{votes}[\hat{y}_k] \nonumber,
\end{align}
where $\text{votes}[y]$ is the total votes for $y$ from classifiers. Note that $H>1$ if we use $f^{\text{MV}}$ and $H=1$ if we use $f^{\text{LOOCV}}$.

\section{Experiment results}\label{sec:experiment_results}

\subsection{Experimental setting}

To obtain stable results, all evaluations are repeated over 10 independent runs. In cases where multiple classifiers yield identical accuracy, the label is assigned randomly between the best classifiers. We conduct the survey on the meta-dataset before training; the results are presented in Table~\ref{tab:pre_evaluation}, showing the distribution of labels $\mathcal{Y}_i$.

\begin{table*}[t]
    \centering
    \caption{Pre-evaluation on meta-dataset.}
    \label{tab:pre_evaluation}
    \begin{tabular}{lll}
        \toprule
        \multicolumn{3}{c}{\textbf{Summary}} \\
        \midrule
        \textbf{Statistic} & $k=1$ & $k>1$ \\
        \midrule
        \#Samples & 200 & 442 \\
        \#Features & 24 & 24 \\
        % \#Circuit classes & 8 & 8 \\
        \midrule
        \multicolumn{3}{c}{\textbf{Frequency of each circuit appearing in $\{\mathcal{Y}_i$\}?
        }} \\
        \midrule
        \textbf{Circuit} & $k=1$ & $k>1$  \\
        \midrule
        HD & 66 (33.0\%) & 89 (20.1\%) \\
        ZFM & 44 (22.0\%) & 89 (20.1\%) \\
        HZY\_CZ & 26 (13.0\%) & 58 (13.1\%) \\
        HERx & 23 (11.5\%) & 85 (19.2\%) \\
        ZZFM & 19 (9.5\%) & 42 (9.5\%) \\
        SRx & 18 (9.0\%) & 74 (16.7\%) \\
        PZFM & 3 (1.5\%) & 3 (0.7\%) \\
        Chebyshev & 1 (0.5\%) & 2 (0.5\%) \\
        YZ\_CX & -- & -- \\
        \bottomrule
    \end{tabular}
\end{table*}

\subsection{Training accuracy}

\begin{sidewaysfigure}
    \centering
    % Điều chỉnh width=1\textwidth để hình ảnh tràn lề đẹp mắt
    \includegraphics[width=1\textwidth]{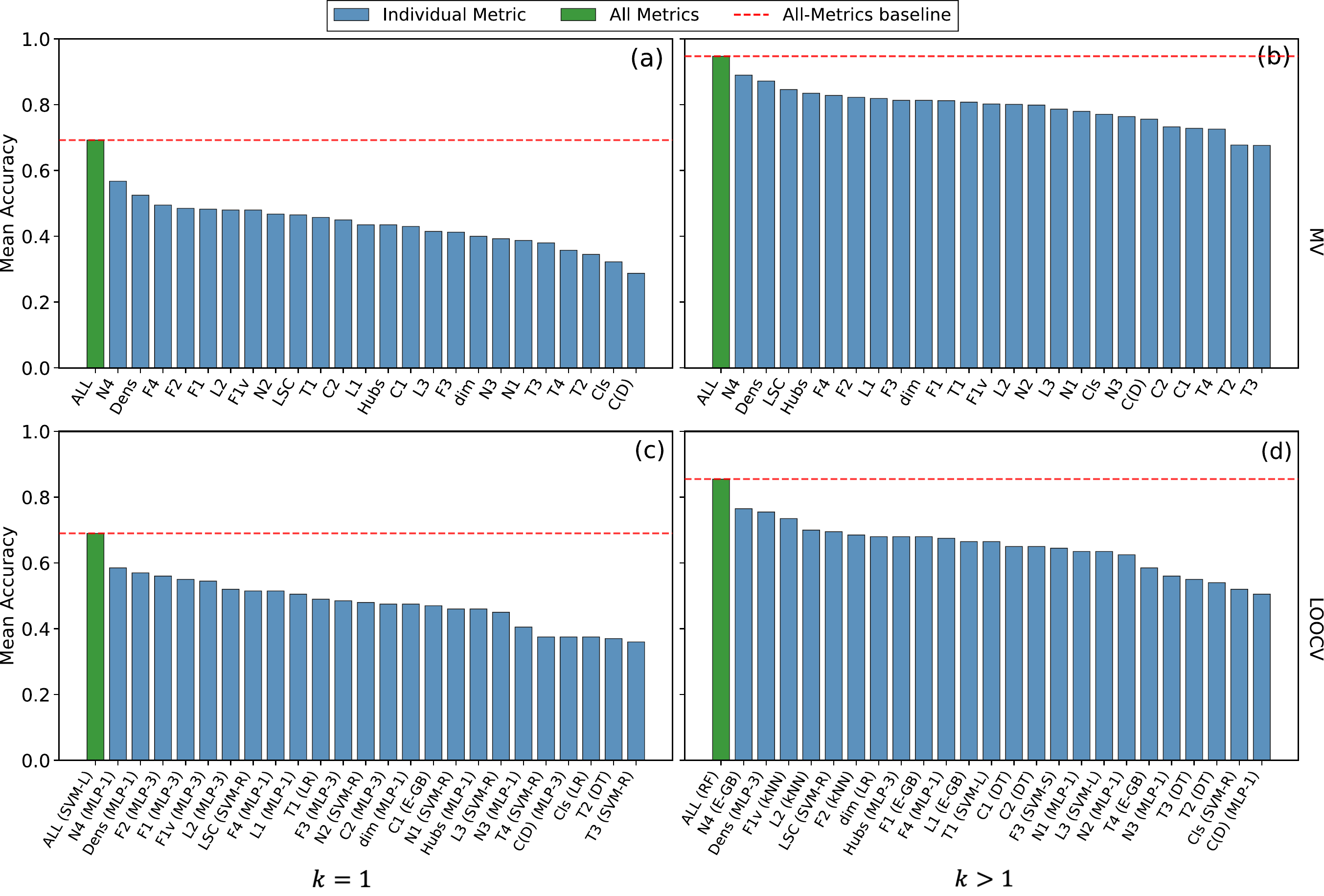}
    \caption{Mean accuracy of \textbf{SINGLE-IN} mode (blue bars) versus $\textbf{ALL-IN}$ mode (the first green bar). In the MV strategy (top), each bar represents the mean accuracy of the classifier ensemble. In the LOOCV strategy (bottom), each bar represents the best feature-classifier configuration identified for that metric. The red dashed line marks the highest mean accuracy. $k>1$ (\textbf{TIED-BEST OUT}) show superiority over $k=1$ (\textbf{SINGLE-BEST OUT}).}
    \label{fig:mean_accuracy}
\end{sidewaysfigure}

Figure~\ref{fig:mean_accuracy} (a)-(b) present the mean accuracy of $f^{\text{MV}}$. The full feature set generally yields same or superior performance compared to individual metrics. In \textbf{ALL-IN} $+$ \textbf{TIED-BEST-OUT} configuration, $f^{\text{MV}}$ achieves the highest mean accuracy compared to others, confirming that complementary information across different complexity measures strengthens prediction.

However, \textbf{ALL-IN} $+$ \textbf{SINGLE-BEST-OUT} configuration's result remains below $0.80$, reflecting the inherent difficulty. Individual metrics show a wide spread around the baseline, and no single metric consistently dominates. Otherwise, consider \textbf{TIED-BEST-OUT} configurations, overall performance improves substantially, with most configurations exceeding $0.80$ accuracy and several approaching $0.90$. The classifier needs only to identify which broad family of encoding circuits is well-suited. As a result, \textbf{TIED-BEST-OUT} is adopted as the primary evaluation framework for subsequent analyses.

Figure~\ref{fig:mean_accuracy} (c)-(d) shows the corresponding LOOCV results. Consistent with MV, \textbf{ALL-IN} $+$ \textbf{SINGLE-BEST-OUT} (SVM-L classifier) achieves 
$\approx 0.69$ and \textbf{ALL-IN} $+$ \textbf{TIED-BEST-OUT} (RF classifier) 
achieves $\approx 0.85$. Top configurations using $N_4$, Density, and $F_2$ metrics achieve around $0.55$--$0.59$ for $k=1$, while most others fall within $0.50$--$0.76$.

\subsection{Evaluate the recommender}

\begin{table}
    \caption{Classification accuracy on new datasets.}
    \label{tab:ground_truth}
    \begin{tabular}{llllllll}
        \toprule
        \textbf{Circuit} & \textbf{Digits\_0v1} & \textbf{Moons\_n35} & \textbf{Circles} & \textbf{Synthetic} & \textbf{GaussianQ} & \textbf{MultiRing} & \textbf{Imbalanced} \\
        \midrule
        YZ\_CX    & 0.4931          & 0.4825          & 0.5525          & 0.5500          & 0.5333          & 0.6633          & 0.8000          \\
        HD        & \textbf{1.0000} & \textbf{0.8600} & \textbf{0.8800} & \textbf{0.8625} & 0.8900          & 0.8333          & \textbf{0.9950} \\
        HZY\_CZ   & 0.9986          & 0.8100          & 0.8625          & 0.8550          & 0.8900          & 0.7167          & \textbf{0.9950} \\
        Chebyshev & 0.5694          & 0.5450          & 0.5450          & 0.5375          & 0.4933          & 0.6700          & 0.7950          \\
        PZFM      & 0.7236          & 0.5700          & 0.5700          & 0.6150          & 0.6167          & 0.6800          & 0.8000          \\
        SRx       & 0.9986          & 0.7775          & 0.8675          & \textbf{0.8625} & 0.9167          & 0.6800          & 0.9800          \\
        HERx      & \textbf{1.0000} & 0.7875          & 0.8525          & \textbf{0.8625} & \textbf{0.9400} & 0.6900          & 0.9900          \\
        ZFM       & 0.9792          & 0.6600          & 0.8525          & 0.8525          & 0.8567          & \textbf{0.9867} & 0.9550          \\
        ZZFM      & 0.8972          & 0.6550          & 0.8525          & 0.8025          & 0.7700          & 0.8267          & 0.9200          \\
        \bottomrule
    \end{tabular}
\end{table}

To evaluate the framework's generalization capability, we use $f$ on new datasets. The recommended encoding circuits' performance is reported in Table~\ref{tab:ground_truth}, the reported accuracy corresponds to the maximum accuracy achieved among the three classifiers (SVC, GPC, KRC). 

From the results, it is evident that several datasets are simultaneously well-suited to multiple encoding circuits. For example, on \texttt{Digits\_0v1}, both \texttt{HighDim} and \texttt{HardwareEfficientRx} achieve perfect accuracy of $1.0000$, while on \texttt{Synthetic}, three encoding circuits - \texttt{HighDim}, \texttt{HardwareEfficientRx}, and \texttt{SeparableRx} - share the same highest score of $0.8625$. This empirical observation reinforces the \textbf{TIED-BEST-OUT} and supports reporting Top-$3$ predictions as the primary evaluation criterion.

Specifically, to assess $f^{\text{MV}}$, we have shown detailed results on four configurations (Table~\ref{tab:MV_prediction_combined}). The \textbf{ALL-IN} mode achieves 42.9\% Top-$1$ accuracy and notably up to 85.7\% Top-$3$ accuracy. By contrast, the \textbf{SINGLE-IN} (using $N_4$ metric) yields only $28.6\%$ Top-$1$ and $71.4\%$ Top-$3$ accuracy, confirming that a single metric is insufficient for reliable encoding circuit selection. It is worth noting that for certain datasets, the Top-$3$ list contains only a single entry. This occurs when all models in the ensemble unanimously vote for the same circuit, leaving no alternative candidates to populate the remaining positions.

\begin{table}[t]
    \centering
    \caption{$f^{\text{MV}}$ prediction. \textbf{SINGLE-IN} mode uses $N_4$ metric.}
    \label{tab:MV_prediction_combined}

        \begin{tabular}{lllll}
            \toprule
            \textbf{\makecell{Mode}} 
            & \textbf{Dataset} 
            & \textbf{Label $\mathcal{Y}_i$} 
            & \textbf{Predicted Top-$1$} 
            & \textbf{Predicted Top-$3$} \\
            \midrule

            \multirow{7}{*}{\textbf{ALL-IN}}
                & Digits\_0v1       & HERx, HD      & SRx (\xmark)  & SRx, HZY, HD (\cmark) \\
                & Moons\_n35        & HD            & HD (\cmark)   & HD, ZFM, ZZFM (\cmark) \\
                & Circles           & HD            & HERx (\xmark) & HERx, SRx, ZFM (\xmark) \\
                & Synthetic         & HERx, HD, SRx & HD (\cmark)   & HD, SRx, HERx (\cmark) \\
                & GaussianQuantiles & HERx          & ZFM (\xmark)  & ZFM, HERx, SRx (\cmark) \\
                & MultiRing         & ZFM           & ZFM (\cmark)  & ZFM (\cmark) \\
                & Imbalanced        & HZY, HD       & ZFM (\xmark)  & ZFM, HZY, SRx (\cmark) \\
            \midrule
                & \multicolumn{2}{r}{\textbf{Final accuracy}} 
                & \textbf{3/7 (42.9\%)} & \textbf{Top-$3$: 6/7 (85.7\%)} \\
            \midrule

            \multirow{7}{*}{\textbf{SINGLE-IN}}
                & Digits\_0v1       & HERx, HD      & SRx (\xmark)  & SRx, HD, HZY (\cmark) \\
                & Moons\_n35        & HD            & HERx (\xmark) & HERx, HD, ZFM (\cmark) \\
                & Circles           & HD            & ZFM (\xmark)  & ZFM, SRx, ZZFM (\xmark) \\
                & Synthetic         & HERx, HD, SRx & HD (\cmark)   & HD, HERx (\cmark) \\
                & GaussianQuantiles & HERx          & ZFM (\xmark)  & ZFM, ZZFM (\xmark) \\
                & MultiRing         & ZFM           & ZFM (\cmark)  & ZFM, Chebyshev (\cmark) \\
                & Imbalanced        & HZY, HD       & HERx (\xmark) & HERx, HD, SRx (\cmark) \\
            \midrule
                & \multicolumn{2}{r}{\textbf{Final accuracy}} & \textbf{2/7 (28.6\%)} & \textbf{5/7 (71.4\%)} \\
            \bottomrule
        \end{tabular}
\end{table}

Table~\ref{tab:LOOCV_prediction_combined} details the per-dataset LOOCV predictions. The \textbf{ALL-IN} rows achieves 71.4\% Top-$1$ and 85.7\% Top-$3$ accuracy, with the only Top-$3$ miss occurring on the \texttt{Circles} dataset, where the ground-truth circuit (\texttt{HD}, accuracy $0.8800$) falls outside this set despite being the best performer, as several competing encoding circuit achieve closely similar accuracy, resulting in a particularly flat performance landscape that challenges the recommender's discrimination ability. In contrast, \textbf{SINGLE-IN} (using $N_4$ metric and E-GB classifier) drops to 28.6\% Top-$1$ and 71.4\% Top-$3$ accuracy, exhibiting a strong bias 
toward predicting ZFM even when it is not the ground-truth best circuit.

\begin{table}[t]
    \centering
    \caption{$f^{\text{LOOCV}}$ prediction in \textbf{ALL-IN} mode (using RF classifier) and \textbf{SINGLE-IN} mode (using E-GB classifier \& $N_4$ metric).}
    \label{tab:LOOCV_prediction_combined}
        \begin{tabular}{lllll}
            \toprule
            \textbf{\makecell{Mode}} 
            & \textbf{Dataset} 
            & \textbf{Label $\mathcal{Y}_i$} 
            & \textbf{Predicted Top-$1$} 
            & \textbf{Predicted Top-$3$} \\
            \midrule

            \multirow{7}{*}{\textbf{ALL-IN}}
                & Digits\_0v1       & HERx, HD      & HD (\cmark)   & HD, HZY, HERx (\cmark) \\
                & Moons\_n35        & HD            & HD (\cmark)   & HD, ZFM, SRx (\cmark) \\
                & Circles           & HD            & SRx (\xmark)  & SRx, HERx, ZFM (\xmark) \\
                & Synthetic         & HERx, HD, SRx & SRx (\cmark)  & SRx, HD, HERx (\cmark) \\
                & GaussianQuantiles & HERx          & SRx (\xmark)  & SRx, ZFM, HERx (\cmark) \\
                & MultiRing         & ZFM           & ZFM (\cmark)  & ZFM, ZZFM, SRx (\cmark) \\
                & Imbalanced        & HZY, HD       & HZY (\cmark)  & HZY, SRx, HERx (\cmark) \\
            \midrule
                & \multicolumn{2}{r}{\textbf{Final accuracy}} 
                & \textbf{5/7 (71.4\%)} 
                & \textbf{6/7 (85.7\%)} \\
            \midrule

            \multirow{7}{*}{\textbf{SINGLE-IN}}
                & Digits\_0v1       & HERx, HD      & SRx (\xmark)  & SRx, HZY, HD (\cmark) \\
                & Moons\_n35        & HD            & HERx (\xmark) & HERx, SRx, HZY (\xmark) \\
                & Circles           & HD            & ZFM (\xmark)  & ZFM, SRx, HERx (\xmark) \\
                & Synthetic         & HERx, HD, SRx & HD (\cmark)   & HD, ZFM, HERx (\cmark) \\
                & GaussianQuantiles & HERx          & ZFM (\xmark)  & ZFM, SRx, HERx (\cmark) \\
                & MultiRing         & ZFM           & ZFM (\cmark)  & ZFM, SRx, HERx (\cmark) \\
                & Imbalanced        & HZY, HD       & HERx (\xmark) & HERx, SRx, HD (\cmark) \\
            \midrule
                & \multicolumn{2}{r}{\textbf{Final accuracy}} 
                & \textbf{2/7 (28.6\%)} 
                & \textbf{5/7 (71.4\%)} \\
            \bottomrule
        \end{tabular}
\end{table}
\subsection{Complexity analysis}

The agreement between two independent evaluation strategies reinforces the core finding: classical data complexity metrics carry sufficient information to find $f$ without any quantum evaluation. The two methods differ in how they aggregate evidence. MV ensembles the predictions of $H$ classifiers into a single consensus vote, which smooths out individual classifier weaknesses and yields stable recommendations. LOOCV, by contrast, evaluates each classifier and maximizes training data utilization by reserving only one sample for testing per fold, providing a per-model performance profile that reveals which models are most reliable for the recommendation task.

From a complexity perspective, the two approaches exhibit different scaling behaviors. Let $N$ denote the number of meta-datasets, $H$ the number of classifiers, and $T(n)$ the training cost of a single classifier on $n$ samples. MV performs $R$ random train/test splits, each requiring training of all $H$ classifiers once; LOOCV requires $N$ leave-one-out iterations per classifier. Then, the ratio between the two is therefore:

\begin{align}
    \frac{\mathcal{C}_{\text{MV}}}{\mathcal{C}_{\text{LOOCV}}} = \frac{\mathcal{O}\bigl(R \cdot H \cdot T(n)\bigr)}{ \mathcal{O}\bigl(H \cdot N \cdot T(n-1)\bigr)} \approx \mathcal{O}(\frac{R}{N})
    \label{eq:complexity_ratio}
\end{align}

For the current setup with $R = 10$, $H = 14$, and $N = 200$, LOOCV is approximately 20 times more expensive than MV in terms of training cost. This difference becomes more pronounced as the size of the dataset grows, making MV the more scalable option for practical deployment, while LOOCV remains valuable as a rigorous validation tool during framework development.

\section{Conclusion and future work}
\label{sec:conclusion}

We presented a meta-learning framework for automated encoding circuit selection in quantum kernel methods, showing that classical data complexity metrics can effectively predict high-performing circuits for an unseen dataset, significantly reducing the need for exhaustive quantum evaluation. The \textbf{TIED-BEST-OUT} task proves consistently more reliable than the \textbf{SINGLE-BEST-OUT} task, reflecting the practical reality that quantum kernel performance landscapes are often flat, a finding that has implications beyond recommendation accuracy, suggesting that practitioners need not commit to a single circuit when several perform equivalently. The Top-$3$ recommendation strategy effectively reduces the quantum evaluation burden from nine candidates to at most three, with no loss of coverage under MV and minimal loss under LOOCV.

Future work may incorporate quantum descriptors such as expressibility, entanglement entropy, and connected correlators~\cite{pere2025datacomplexitythresholdclassical}.  We will conduct the two-stage pipeline to refine the initial Top-$k$ candidates. Furthermore, the current framework should be extended to regression and combinatorial optimization problems would broaden its applicability to a wider range of quantum computing workloads. Rather than selecting from  existing encoding circuits, a natural progression is to generate problem-specific circuit architectures directly from $\mathcal{X}$, moving from recommender toward designer~\cite{neto2025datacomplexitymeasuresquantum}. Scaling the meta-dataset beyond $200$ instances and evaluating under noise models are additional directions to strengthen generalization and practical applicability.

\begin{figure*}[http]
    \centering
    \includegraphics[width=1\textwidth]{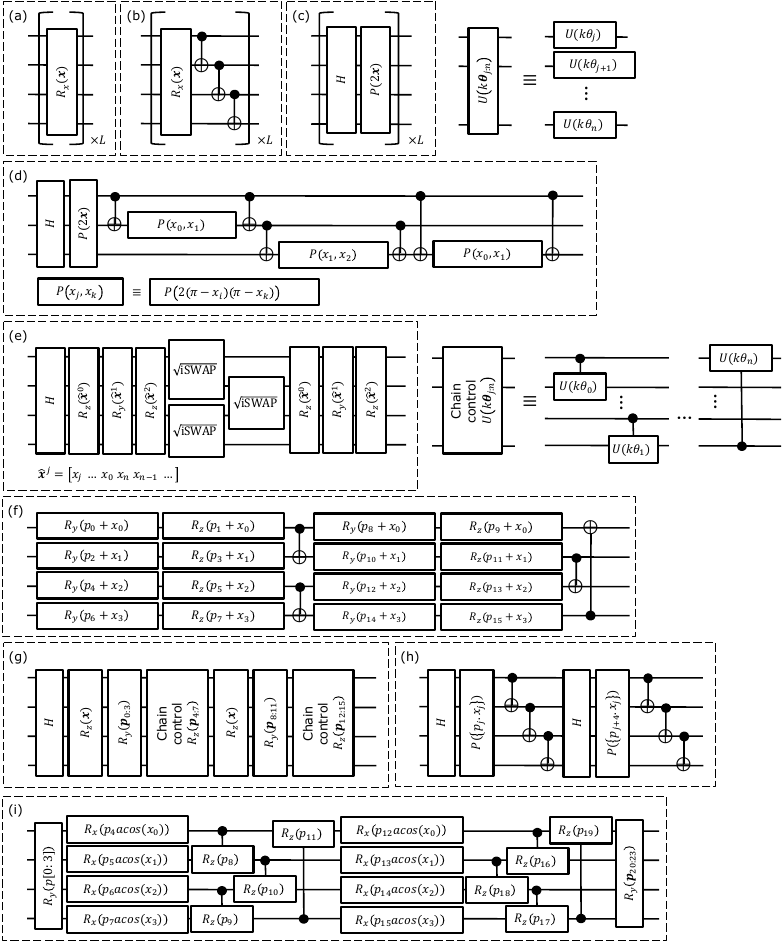}
    \caption{Diagrams of the nine encoding circuits: (a)~SRx\cite{1644040}, (b)~HERx \cite{Thanasilp2024}, (c)~ZFM\cite{qiskit2024}, (d)~ZZFM\cite{qiskit2024}, (e)~HD\cite{Peters2021}, (f)~YZ\_CX \cite{Haug_2023}, (g)~HZY\_CZ\cite{Hubregtsen2021}, (h)~PZFM \cite{qiskit2024}, (i)~(h)~Chebyshev \cite{Kreplin2024reductionoffinite}. These circuits encode the input $\bm x=[x_0\;\ldots,x_n]$.}
    \label{fig:circuits}
\end{figure*}

\section*{Acknowledgment}

L. N. T is funded by the National Foundation for Science and Technology Development (NAFOSTED) Grant Number 103.01-2024.06. L. B. H is funded by the Tohoku Initiative for
Fostering Global Researchers for Interdisciplinary Sciences (TI-FRIS) of MEXT’s Strategic Professional Development Program for Young Researchers.

\section*{Data and code availability}

Data are available from the corresponding authors upon reasonable request. The code is available
at \href{https://github.com/tungduy1704/Meta-learning_for_Ansatz_Selection}{Github}.

\begin{appendices}\label{sec:appendix}

\section{Classical descriptors}

The 24 complexity metrics are used as features given in Table~\ref{tab:complexity_metrics}; including 22 classification measures from problexity~\cite{komorniczak2023problexity, lorena2018complex}, together with Intrinsic Dimension and Kolmogorov Complexity from Qsun~\cite{Nguyen_2022}.

\begin{table*}[t]
    \centering
    \caption{List of classical descriptors.}
    \label{tab:complexity_metrics}
    \renewcommand{\arraystretch}{1.3}
    \begin{tabularx}{0.99\linewidth}{|p{2.8cm}|p{2.3cm}|X|}
        \hline
        \textbf{Category} & \textbf{Metrics} & \textbf{Description} \\
        \hline

        Feature-based 
        & $F_1$\cite{22}, $F_{1v}$\cite{938265}, $F_2$\cite{sss}, $F_3,F_4$\cite{lorena2018complex}
        & Characterize how informative features are to separate classes via range, spread, and overlap. \\
        \hline

        Linearity 
        & $L_1, L_2$ \cite{1687347}, $L_3$\cite{lorena2018complex, 547429} 
        & Quantify whether the classes can be linearly separated, by measuring the error and nonlinearity of linear classifiers fitted to the data. \\
        \hline

        Neighborhood 
        & $N_1, N_2, N_3, N_4$ \cite{lorena2018complex}, $T_1$\cite{990132,lorena2018complex}, LSC\cite{6823733,lorena2018complex} 
        & Characterize the presence and density of the same or different classes in local neighborhoods, using nearest-neighbor distances, graph-based boundary analysis, and local set cardinality. \\
        \hline

        Network 
        & Density, ClsCoef, Hubs \cite{lorena2018complex}
        & Extract structural information from the dataset by modeling it as a graph, capturing edge density, clustering tendency, and hub connectivity among same-class instances. \\
        \hline

        Dimensionality 
        & $T_2, T_3, T_4$  \cite{LORENA201233,lorena2018complex}
        & Estimate sparsity and intrinsic dimensionality (e.g., PCA-based). \\
        \hline

        Class Balance 
        & $C_1$\cite{LORENA201233,lorena2018complex}, $C_2$ \cite{10.1007/978-3-642-17508-4_9,lorena2018complex}
        & Measure the degree of imbalance in class distributions using entropy and class proportion ratios. \\
        \hline

        Intrinsic dimension
        & $\dim_{\text{eff}}$  \cite{bishop2016pattern} 
        & Effective intrinsic dimensionality from covariance eigenspectrum. \\
        \hline

        Kolmogorov complexity 
        & $C(D)$ \cite{pere2025datacomplexitythresholdclassical} 
        & Kolmogorov complexity approximated by the compression ratio of the dataset. \\
        \hline

    \end{tabularx}
\end{table*}

\section{Encoding circuits}
\label{app:circuit}

In the following, we provide circuit diagrams used in this study, as shown in Figure~\ref{fig:circuits}. Features are denoted with a feature vector $\bm x$, while variationally trainable parameters ($\bm\theta$) are labeled $\bm p$. Table~\ref{tab:ansatz_properties} summarizes the structural properties of each circuit, including the number of parameters (\#Params), number of gates (\#Gates), circuit depth, and the type of two-qubit entangling gate used.

\begin{table*}[t]
    \centering
    \caption{Properties of the encoding circuits and their abbreviation used in this study ($L=2$ layers).}
    \label{tab:ansatz_properties}
    \begin{tabular}{p{3.3cm}p{1.5cm}c c c p{3cm}}
        \toprule
        \textbf{Circuit} 
        & \textbf{Abbr.}
        & \textbf{\#Params}
        & \textbf{\#Gates}
        & \textbf{Depth}
        & \textbf{2-qubit gate} \\
        \midrule
        SeparableRx            & SRx        & 0  & 8  & 2  & - \\
        HardwareEfficientRx    & HERx       & 0  & 14 & 8  & CX \\
        ZFeatureMap            & ZFM        & 0  & 16 & 4  & - \\
        ZZFeatureMap           & ZZFM       & 0  & 34 & 22 & CX \\
        HighDim                & HD         & 0  & 31 & 9  & $\sqrt{i\text{SWAP}}$ \\
        YZ\_CX                 & -          & 16 & 20 & 6  & CX \\
        HZY\_CZ                & -          & 16 & 28 & 13 & CRZ \\
        ParamZFeatureMap       & PZFM       & 8  & 22 & 10 & CX \\
        Chebyshev              & -          & 24 & 24 & 10 & CRZ \\ 
        \bottomrule
    \end{tabular}
\end{table*}

\section{Classifiers}
\label{classifiers}
We employ 14 different classifiers for our recommendation system, summarized in Table~\ref{tab:classifiers}. These classifiers are drawn from diverse categories of supervised learning algorithms, ensuring diversity in inductive biases. This diversity is particularly important for MV strategy, where ensemble consensus benefits from classifiers that make independent errors, and also broadens the search space for LOOCV.

\begin{table*}[t]
    \centering
    \caption{Classifiers used in the experiments.}
    \label{tab:classifiers}
        \begin{tabular}{>{\raggedright\arraybackslash}p{3cm}p{4cm}p{2cm}p{3.5cm}}
            \toprule
            \textbf{Category} & \textbf{Classifier} & \textbf{Abbr.} & \textbf{Parameters} \\
            \midrule
            \multirow{2}{*}{\makecell[l]{Tree-based}} & Decision Tree & DT & max\_depth=None \\
            & Random Forest & RF & n\_estimators=10 \\
            \midrule
            \multirow{3}{*}{Ensemble} & Gradient Boosting & E-GB & n\_estimators=100 \\
            & AdaBoost & AB & n\_estimators=50 \\
            & Bagging & Bg & n\_estimators=10 \\
            \midrule
            \multirow{3}{*}{SVM} & SVM-Linear & SVM-L & kernel=`linear' \\
            & SVM-RBF & SVM-R & kernel=`rbf', C=1.0 \\
            & SVM-Sigmoid & SVM-S & kernel=`sigmoid' \\
            \midrule
            \multirow{2}{=}{Neural network} & MLP (500) & MLP-1 & hidden=(500,) \\
            & MLP (100-100-100) & MLP-3 & hidden=(100,100,100) \\
            \midrule
            \multirow{2}{=}{Instance-based} & $k$-NN & KNN & n\_neighbors=5 \\
            & Nearest Centroid & NC & metric=`euclidean' \\
            \midrule
            \multirow{2}{*}{Probabilistic} & Naive Bayes & NB & Gaussian \\
            & Logistic Regression & LR & max\_iter=1000 \\
            \bottomrule
        \end{tabular}
\end{table*}

\end{appendices}

\bibliography{citation}% 

@misc{neto2025datacomplexitymeasuresquantum,
      title={Data Complexity Measures for Quantum Circuits Architecture Recommendation}, 
      author={Fernando M de Paula Neto},
      year={2025},
      archivePrefix={arXiv},
      primaryClass={cs.LG},
      doi={https://doi.org/10.48550/arXiv.2502.15129}
}

@article{10812182,
  author={Incudini, Massimiliano and Bosco, Daniele Lizzio and Martini, Francesco and Grossi, Michele and Serra, Giuseppe and Pierro, Alessandra Di},
  journal={IEEE Transactions on Emerging Topics in Computational Intelligence}, 
  title={Automatic and Effective Discovery of Quantum Kernels}, 
  year={2024},
  volume={},
  number={},
  pages={1-10},
  doi={10.1109/TETCI.2024.3499993}}

@article{rspa,
    author = {Ciliberto, Carlo and Herbster, Mark and Ialongo, Alessandro Davide and Pontil, Massimiliano and Rocchetto, Andrea and Severini, Simone and Wossnig, Leonard},
    title = {Quantum machine learning: a classical perspective},
    journal = {Proceedings of the Royal Society A: Mathematical, Physical and Engineering Sciences},
    volume = {474},
    number = {2209},
    pages = {20170551},
    year = {2018},
    month = {01},
    issn = {1364-5021},
    doi = {10.1098/rspa.2017.0551}
}

@misc{adcock2015advancesquantummachinelearning,
      title={Advances in quantum machine learning}, 
      author={Jeremy Adcock and Euan Allen and Matthew Day and Stefan Frick and Janna Hinchliff and Mack Johnson and Sam Morley-Short and Sam Pallister and Alasdair Price and Stasja Stanisic},
      year={2015},
      archivePrefix={arXiv},
      primaryClass={quant-ph},
      doi ={https://doi.org/10.48550/arXiv.1512.02900}, 
}

@Article{Biamonte2017,
author={Biamonte, Jacob
and Wittek, Peter
and Pancotti, Nicola
and Rebentrost, Patrick
and Wiebe, Nathan
and Lloyd, Seth},
title={Quantum machine learning},
journal={Nature},
year={2017},
month={Sep},
day={01},
volume={549},
number={7671},
pages={195-202},
issn={1476-4687},
doi={10.1038/nature23474}
}

@article{doi:10.1137/S0097539795293172,
author = {Shor, Peter W.},
title = {Polynomial-Time Algorithms for Prime Factorization and Discrete Logarithms on a Quantum Computer},
journal = {SIAM Journal on Computing},
volume = {26},
number = {5},
pages = {1484-1509},
year = {1997},
doi = {10.1137/S0097539795293172},
}

@inproceedings{10.1145/237814.237866,
author = {Grover, Lov K.},
title = {A fast quantum mechanical algorithm for database search},
year = {1996},
isbn = {0897917855},
publisher = {Association for Computing Machinery},
address = {New York, NY, USA},
doi = {10.1145/237814.237866},
booktitle = {Proceedings of the Twenty-Eighth Annual ACM Symposium on Theory of Computing},
pages = {212–219},
numpages = {8},
location = {Philadelphia, Pennsylvania, USA},
series = {STOC '96}
}

@Article{Cortes1995,
author={Cortes, Corinna
and Vapnik, Vladimir},
title={Support-vector networks},
journal={Machine Learning},
year={1995},
month={Sep},
day={01},
volume={20},
number={3},
pages={273-297},
issn={1573-0565},
doi={10.1007/BF00994018}
}

@article{doi:10.1137/16M1080173,
author = {Bottou, L\'{e}on and Curtis, Frank E. and Nocedal, Jorge},
title = {Optimization Methods for Large-Scale Machine Learning},
journal = {SIAM Review},
volume = {60},
number = {2},
pages = {223-311},
year = {2018},
doi = {10.1137/16M1080173},
}

@Article{Liu2021,
author={Liu, Yunchao
and Arunachalam, Srinivasan
and Temme, Kristan},
title={A rigorous and robust quantum speed-up in supervised machine learning},
journal={Nature Physics},
year={2021},
month={Sep},
day={01},
volume={17},
number={9},
pages={1013-1017},
issn={1745-2481},
doi={10.1038/s41567-021-01287-z}
}

@Article{Peruzzo2014,
author={Peruzzo, Alberto
and McClean, Jarrod
and Shadbolt, Peter
and Yung, Man-Hong
and Zhou, Xiao-Qi
and Love, Peter J.
and Aspuru-Guzik, Al{\'a}n
and O'Brien, Jeremy L.},
title={A variational eigenvalue solver on a photonic quantum processor},
journal={Nature Communications},
year={2014},
month={Jul},
day={23},
volume={5},
number={1},
pages={4213},
issn={2041-1723},
doi={10.1038/ncomms5213}
}

@Article{Larocca2025,
author={Larocca, Mart{\'i}n
and Thanasilp, Supanut
and Wang, Samson
and Sharma, Kunal
and Biamonte, Jacob
and Coles, Patrick J.
and Cincio, Lukasz
and McClean, Jarrod R.
and Holmes, Zo{\"e}
and Cerezo, M.},
title={Barren plateaus in variational quantum computing},
journal={Nature Reviews Physics},
year={2025},
month={Apr},
day={01},
volume={7},
number={4},
pages={174-189},
issn={2522-5820},
doi={10.1038/s42254-025-00813-9}
}

@Article{Cerezo2021,
author={Cerezo, M.
and Arrasmith, Andrew
and Babbush, Ryan
and Benjamin, Simon C.
and Endo, Suguru
and Fujii, Keisuke
and McClean, Jarrod R.
and Mitarai, Kosuke
and Yuan, Xiao
and Cincio, Lukasz
and Coles, Patrick J.},
title={Variational quantum algorithms},
journal={Nature Reviews Physics},
year={2021},
month={Sep},
day={01},
volume={3},
number={9},
pages={625-644},
issn={2522-5820},
doi={10.1038/s42254-021-00348-9}
}

@Article{Huang2021,
author={Huang, Hsin-Yuan
and Broughton, Michael
and Mohseni, Masoud
and Babbush, Ryan
and Boixo, Sergio
and Neven, Hartmut
and McClean, Jarrod R.},
title={Power of data in quantum machine learning},
journal={Nature Communications},
year={2021},
month={May},
day={11},
volume={12},
number={1},
pages={2631},
issn={2041-1723},
doi={10.1038/s41467-021-22539-9}
}

@article{komorniczak2023problexity,
title = {problexity-An open-source Python library for supervised learning problem complexity assessment},
journal = {Neurocomputing},
volume = {521},
pages = {126-136},
year = {2023},
issn = {0925-2312},
doi = {https://doi.org/10.1016/j.neucom.2022.11.056},
author = {Joanna Komorniczak and Paweł Ksieniewicz},
}

@article{lorena2018complex,
author = {Lorena, Ana C. and Garcia, Lu\'{\i}s P. F. and Lehmann, Jens and Souto, Marcilio C. P. and Ho, Tin Kam},
title = {How Complex Is Your Classification Problem? A Survey on Measuring Classification Complexity},
year = {2019},
issue_date = {September 2020},
publisher = {Association for Computing Machinery},
address = {New York, NY, USA},
volume = {52},
number = {5},
issn = {0360-0300},
doi = {10.1145/3347711},
month = sep,
articleno = {107},
numpages = {34}
}

@article{Nguyen_2022,
doi = {10.1088/2632-2153/ac5997},
year = {2022},
month = {mar},
publisher = {IOP Publishing},
volume = {3},
number = {1},
pages = {015034},
author = {Nguyen, Quoc Chuong and Ho, Le Bin and Nguyen Tran, Lan and Nguyen, Hung Q},
title = {Qsun: an open-source platform towards practical quantum machine learning applications},
journal = {Machine Learning: Science and Technology}
}

@article{PhysRevLett.122.040504,
  title = {Quantum Machine Learning in Feature Hilbert Spaces},
  author = {Schuld, Maria and Killoran, Nathan},
  journal = {Phys. Rev. Lett.},
  volume = {122},
  issue = {4},
  pages = {040504},
  numpages = {6},
  year = {2019},
  month = {Feb},
  publisher = {American Physical Society},
  doi = {10.1103/PhysRevLett.122.040504}
}

@misc{schuld2021supervisedquantummachinelearning,
      title={Supervised quantum machine learning models are kernel methods}, 
      author={Maria Schuld},
      year={2021},
      archivePrefix={arXiv},
      primaryClass={quant-ph},
      doi={https://doi.org/10.48550/arXiv.2101.11020}, 
}

@article{22,
  title={A review of data complexity measures and their applicability to pattern classification problems},
  author={Sotoca, Jos{\'e} M and S{\'a}nchez, Jos{\'e} Salvador and Mollineda, Ram{\'o}n A},
  journal={Actas del III Taller Nacional de Mineria de Datos y Aprendizaje},
  volume={1},
  pages={18--44},
  year={2005},
  publisher={TAMIDA},
  doi={ }
}

@ARTICLE{990132,
  author={Tin Kam Ho and Basu, M.},
  journal={IEEE Transactions on Pattern Analysis and Machine Intelligence}, 
  title={Complexity measures of supervised classification problems}, 
  year={2002},
  volume={24},
  number={3},
  pages={289-300},
  doi={10.1109/34.990132}}

@misc{pere2025datacomplexitythresholdclassical,
      title={Data Complexity: a threshold between Classical and Quantum Machine Learning -- Part I}, 
      author={Christophe Pere},
      year={2025},
      archivePrefix={arXiv},
      primaryClass={quant-ph},
      doi={https://doi.org/10.48550/arXiv.2509.16410}
}

@article{scikit-learn,
  title={Scikit-learn: Machine Learning in {P}ython},
  author={Pedregosa, F. et. al},
  journal={Journal of Machine Learning Research},
  volume={12},
  pages={2825--2830},
  year={2011},
  doi={10.5555/1953048.2078195}
}

@ARTICLE{938265,
  author={Malina, W.},
  journal={IEEE Transactions on Systems, Man, and Cybernetics, Part B (Cybernetics)}, 
  title={Two-parameter Fisher criterion}, 
  year={2001},
  volume={31},
  number={4},
  pages={629-636},
  doi={10.1109/3477.938265}}

@InProceedings{sss,
author="Mollineda, Ram{\'o}n A.
and S{\'a}nchez, J. Salvador
and Sotoca, Jos{\'e} M.",
title="Data Characterization for Effective Prototype Selection",
booktitle="Pattern Recognition and Image Analysis",
year="2005",
publisher="Springer Berlin Heidelberg",
address="Berlin, Heidelberg",
pages="27--34",
doi={10.1007/11492542_4}
}

@ARTICLE{1687347,
  author={Smith, F.W.},
  journal={IEEE Transactions on Computers}, 
  title={Pattern Classifier Design by Linear Programming}, 
  year={1968},
  volume={C-17},
  number={4},
  pages={367-372},
  doi={10.1109/TC.1968.229395}}

@INPROCEEDINGS {547429,
author = { Hoekstra, A. and Duin, R. P. W. },
booktitle = { Pattern Recognition, International Conference on },
title = {{ On the Nonlinearity of Pattern Classifiers }},
year = {1996},
volume = {4},
ISSN = {1051-4651},
pages = {271},
doi = {10.1109/ICPR.1996.547429},
publisher = {IEEE Computer Society},
address = {Los Alamitos, CA, USA},
month =Aug}

@ARTICLE{6823733,
  author={Leyva, Enrique and González, Antonio and Pérez, Raúl},
  journal={IEEE Transactions on Knowledge and Data Engineering}, 
  title={A Set of Complexity Measures Designed for Applying Meta-Learning to Instance Selection}, 
  year={2015},
  volume={27},
  number={2},
  pages={354-367},
  doi={10.1109/TKDE.2014.2327034}}

@article{LORENA201233,
title = {Analysis of complexity indices for classification problems: Cancer gene expression data},
journal = {Neurocomputing},
volume = {75},
number = {1},
pages = {33-42},
year = {2012},
note = {Brazilian Symposium on Neural Networks (SBRN 2010) International Conference on Hybrid Artificial Intelligence Systems (HAIS 2010)},
issn = {0925-2312},
doi = {https://doi.org/10.1016/j.neucom.2011.03.054},
author = {Ana C. Lorena and Ivan G. Costa and Newton Spolaôr and Marcilio C.P. {de Souto}}
}

@InProceedings{10.1007/978-3-642-17508-4_9,
author="Tanwani, Ajay Kumar
and Farooq, Muddassar",
title="Classification Potential vs. Classification Accuracy: A Comprehensive Study of Evolutionary Algorithms with Biomedical Datasets",
booktitle="Learning Classifier Systems",
year="2010",
publisher="Springer Berlin Heidelberg",
address="Berlin, Heidelberg",
pages="127--144",
isbn="978-3-642-17508-4",
doi={10.1007/978-3-642-17508-4_9}
}

@book{bishop2016pattern,
  title={Pattern Recognition and Machine Learning},
  author={Bishop, C.M.},
  isbn={9781493938438},
  lccn={2006922522},
  series={Information Science and Statistics},
  year={2006},
  publisher={Springer},
  address = {New York}
}

@misc{banknote_authentication_267,
  author       = {Lohweg, Volker},
  title        = {{Banknote Authentication}},
  year         = {2012},
  howpublished = {UCI Machine Learning Repository},
  doi         = {https://doi.org/10.24432/C55P57}
}

@misc{habermans_survival_43,
  author       = {Haberman, S.},
  title        = {{Haberman's Survival}},
  year         = {1976},
  howpublished = {UCI Machine Learning Repository},
  doi         = {https://doi.org/10.24432/C5XK51}
}

@inproceedings{Smith1988UsingTA,
  title     = {Using the ADAP Learning Algorithm to Forecast the Onset of Diabetes Mellitus},
  author    = {Smith, Jack W and Everhart, J E and Dickson, W C and Knowler, W C and Johannes, R S},
  booktitle = {Proceedings of the Annual Symposium on Computer Application in Medical Care},
  year      = {1988},
  month     = {11},
  pages     = {261--265},
  note      = {Published November 9, 1988}
}

@article{Haug_2023,
doi = {10.1088/2632-2153/acb0b4},
year = {2023},
month = {jan},
publisher = {IOP Publishing},
volume = {4},
number = {1},
pages = {015005},
author = {Haug, Tobias and Self, Chris N and Kim, M S},
title = {Quantum machine learning of large datasets using randomized measurements},
journal = {Machine Learning: Science and Technology}
}

@Article{Peters2021,
author={Peters, Evan
and Caldeira, Jo{\~a}o
and Ho, Alan
and Leichenauer, Stefan
and Mohseni, Masoud
and Neven, Hartmut
and Spentzouris, Panagiotis
and Strain, Doug
and Perdue, Gabriel N.},
title={Machine learning of high dimensional data on a noisy quantum processor},
journal={npj Quantum Information},
year={2021},
month={Nov},
day={11},
volume={7},
number={1},
pages={161},
issn={2056-6387},
doi={10.1038/s41534-021-00498-9}
}

@Article{Hubregtsen2021,
author={Hubregtsen, Thomas
and Pichlmeier, Josef
and Stecher, Patrick
and Bertels, Koen},
title={Evaluation of parameterized quantum circuits: on the relation between classification accuracy, expressibility, and entangling capability},
journal={Quantum Machine Intelligence},
year={2021},
month={Mar},
day={11},
volume={3},
number={1},
pages={9},
issn={2524-4914},
doi={10.1007/s42484-021-00038-w}
}

@Article{Abbas2021,
author={Abbas, Amira
and Sutter, David
and Zoufal, Christa
and Lucchi, Aurelien
and Figalli, Alessio
and Woerner, Stefan},
title={The power of quantum neural networks},
journal={Nature Computational Science},
year={2021},
month={Jun},
day={01},
volume={1},
number={6},
pages={403-409},
issn={2662-8457},
doi={10.1038/s43588-021-00084-1}
}

@Article{Pan2023,
author={Pan, Xiaoxuan
and Lu, Zhide
and Wang, Weiting
and Hua, Ziyue
and Xu, Yifang
and Li, Weikang
and Cai, Weizhou
and Li, Xuegang
and Wang, Haiyan
and Song, Yi-Pu
and Zou, Chang-Ling
and Deng, Dong-Ling
and Sun, Luyan},
title={Deep quantum neural networks on a superconducting processor},
journal={Nature Communications},
year={2023},
month={Jul},
day={06},
volume={14},
number={1},
pages={4006},
issn={2041-1723},
doi={10.1038/s41467-023-39785-8}
}

@article{10.1145/3529756,
author = {Massoli, Fabio Valerio and Vadicamo, Lucia and Amato, Giuseppe and Falchi, Fabrizio},
title = {A Leap among Quantum Computing and Quantum Neural Networks: A Survey},
year = {2022},
issue_date = {May 2023},
publisher = {Association for Computing Machinery},
address = {New York, NY, USA},
volume = {55},
number = {5},
issn = {0360-0300},
doi = {10.1145/3529756},
journal = {ACM Comput. Surv.},
month = dec,
articleno = {98},
numpages = {37}
}

@article{10.1098/rspa.1909.0075,
    author = {Mercer, James},
    title = {Functions of positive and negative type, and their connection with the theory of integral equations},
    journal = {Proceedings of the Royal Society of London. Series A, Containing Papers of a Mathematical and Physical Character},
    volume = {83},
    number = {559},
    pages = {69-70},
    year = {1909},
    month = {11},    
    issn = {0950-1207},
    doi = {10.1098/rspa.1909.0075},
}

@inproceedings{10.1145/130385.130401,
author = {Boser, Bernhard E. and Guyon, Isabelle M. and Vapnik, Vladimir N.},
title = {A training algorithm for optimal margin classifiers},
year = {1992},
isbn = {089791497X},
publisher = {Association for Computing Machinery},
address = {New York, NY, USA},
doi = {10.1145/130385.130401},
booktitle = {Proceedings of the Fifth Annual Workshop on Computational Learning Theory},
pages = {144–152},
numpages = {9},
location = {Pittsburgh, Pennsylvania, USA},
series = {COLT '92}
}

@article{PhysRevA.106.042431,
  title = {Training quantum embedding kernels on near-term quantum computers},
  author = {Hubregtsen, Thomas and Wierichs, David and Gil-Fuster, Elies and Derks, Peter-Jan H. S. and Faehrmann, Paul K. and Meyer, Johannes Jakob},
  journal = {Phys. Rev. A},
  volume = {106},
  issue = {4},
  pages = {042431},
  numpages = {18},
  year = {2022},
  month = {Oct},
  publisher = {American Physical Society},
  doi = {10.1103/PhysRevA.106.042431},
}

@inproceedings{10.5555/1643031.1643047,
author = {Kohavi, Ron},
title = {A study of cross-validation and bootstrap for accuracy estimation and model selection},
year = {1995},
isbn = {1558603638},
publisher = {Morgan Kaufmann Publishers Inc.},
address = {San Francisco, CA, USA},
booktitle = {Proceedings of the 14th International Joint Conference on Artificial Intelligence - Volume 2},
pages = {1137–1143},
numpages = {7},
location = {Montreal, Quebec, Canada},
doi={10.5555/1643031.1643047},
series = {IJCAI'95}
}

@book{10.7551/mitpress/3206.001.0001,
    author = {Rasmussen, Carl Edward and Williams, Christopher K. I.},
    title = {Gaussian Processes for Machine Learning},
    publisher = {The MIT Press},
    year = {2005},
    month = {11},
    isbn = {9780262256834},
    doi = {10.7551/mitpress/3206.001.0001},
    address = {Cambridge, MA},
    pages={1-244}
}

@ARTICLE{667881,
  author={Kittler, J. and Hatef, M. and Duin, R.P.W. and Matas, J.},
  journal={IEEE Transactions on Pattern Analysis and Machine Intelligence}, 
  title={On combining classifiers}, 
  year={1998},
  volume={20},
  number={3},
  pages={226-239},
  doi={10.1109/34.667881}}

@inproceedings{10.5555/645527.657464,
author = {Saunders, Craig and Gammerman, Alexander and Vovk, Volodya},
title = {Ridge Regression Learning Algorithm in Dual Variables},
year = {1998},
isbn = {1558605568},
publisher = {Morgan Kaufmann Publishers Inc.},
address = {San Francisco, CA, USA},
booktitle = {Proceedings of the Fifteenth International Conference on Machine Learning},
pages = {515–521},
numpages = {7},
series = {ICML '98},
doi={10.5555/645527.657464}
}

@article{Kreplin2024reductionoffinite,
  doi = {10.22331/q-2024-06-25-1385},
  title = {Reduction of finite sampling noise in quantum neural networks},
  author = {Kreplin, David A. and Roth, Marco},
  journal = {{Quantum}},
  issn = {2521-327X},
  publisher = {{Verein zur F{\"{o}}rderung des Open Access Publizierens in den Quantenwissenschaften}},
  volume = {8},
  pages = {1385},
  month = jun,
  year = {2024}
}

@article{1644040,
  author = {Abdulkadir Canatar and Evan Peters and Cengiz Pehlevan and Stefan M. Wild and Ruslan Shaydulin},
  title = {Bandwidth Enables Generalization in Quantum Kernel Models},
  year = {2023},
  journal = {Transactions on Machine Learning Research},
  language = {eng},
  doi={https://doi.org/10.48550/arXiv.2206.06686}
}

@Article{Thanasilp2024,
author={Thanasilp, Supanut
and Wang, Samson
and Cerezo, M.
and Holmes, Zo{\"e}},
title={Exponential concentration in quantum kernel methods},
journal={Nature Communications},
year={2024},
month={Jun},
day={18},
volume={15},
number={1},
pages={5200},
issn={2041-1723},
doi={10.1038/s41467-024-49287-w},
}

@misc{qiskit2024,
      title={Quantum computing with {Q}iskit},
      author={Javadi-Abhari, Ali and Treinish, Matthew and Krsulich, Kevin and Wood, Christopher J. and Lishman, Jake and Gacon, Julien and Martiel, Simon and Nation, Paul D. and Bishop, Lev S. and Cross, Andrew W. and Johnson, Blake R. and Gambetta, Jay M.},
      year={2024},
      doi={10.48550/arXiv.2405.08810},
      archivePrefix={arXiv},
      primaryClass={quant-ph}
}

\end{document}